\documentclass[preprint]{aastex}

\newcommand{\name}{WR~142b}

\slugcomment{Submitted to The Astronomical Journal, Nov 2011}

\shorttitle{Discovery of a Wolf-Rayet Star}
\shortauthors{Littlefield et al.}

\begin{document}


\title{Discovery of a Wolf-Rayet Star Through Detection of its Photometric Variability}


\author{Colin Littlefield$^1$, Peter Garnavich$^2$, G. H. ``Howie'' Marion$^3$,
J\'ozsef Vink\'o$^{4,5}$, Colin McClelland$^2$, Terrence Rettig$^2$, J. Craig Wheeler$^5$}
\altaffiltext{1}{Law School, University of Notre Dame, Notre Dame IN, 46556}
\altaffiltext{2}{Physics Department, University of Notre Dame, Notre Dame IN, 46556}
\altaffiltext{3}{Harvard-Smithsonian Center for Astrophysics, Cambridge, MA 02138}
\altaffiltext{4}{Department of Optics, University of Szeged, Hungary}
\altaffiltext{5}{Astronomy Department, University of Texas, Austin, TX 78712}








\begin{abstract}

We report the serendipitous discovery of a heavily reddened
Wolf-Rayet star that we name \name. While photometrically monitoring a cataclysmic
variable, we detected weak variability in a nearby field star. Low-resolution
spectroscopy revealed a strong emission line at 7100~\AA , suggesting
an unusual object and prompting further study.  A spectrum
taken with the Hobby-Eberly Telescope confirms strong He~II emission
and a N~IV 7112~\AA\ line consistent with a nitrogen-rich Wolf-Rayet
star of spectral class WN6. Analysis of the He~II line strengths reveals
no detectable hydrogen in \name . A blue-sensitive
spectrum obtained with the Large Binocular Telescope shows no evidence
for a hot companion star. The continuum shape and emission line
ratios imply a reddening of $E(B-V)=2.2$ to $2.5$ mag. If not for the
dust extinction, this new Wolf-Rayet star could be visible to the naked eye.

\end{abstract}





\section{Introduction}

Wolf-Rayet (WR) stars have been enigmatic objects
ever since the discovery of the first three WR stars in 1867
\citep{wolf-rayet}. Their spectra
show strong, Doppler-broadened emission lines, primarily from helium
and either carbon or nitrogen of various ionization states. The
nitrogen-rich variety, which are classified as WN stars, outnumber 
the carbon-rich WC stars in current surveys of Galactic WR stars
\citep{hucht01}. \citet{larmers91} conclusively showed 
that WR stars are evolved, high-surface-temperature stars that 
have shed their envelopes via their strong stellar winds. As WR stars lose 
mass, they expose elements created via hydrogen fusion and, eventually, 
those from helium fusion as well \citep{conti76}. They often have masses 
between 10-25 M$_\odot$ and even as high as 80 M$_\odot$ in some cases 
\citep{crowther07}. WR stars are destined to end their lives as type Ib/c
supernovae and possibly gamma-ray bursts (e.g., \citet{smartt09}). 

WR stars are rare, and only about 400 are known in the Galaxy \citep{lopes10}. Because they
are young stars close to the Galactic plane, WR stars are often obscured by dust. Most
recent searches have used infrared wavelengths to avoid dust extinction and
find new members of the WR family (e.g., \citet{wachter10}).

Here, we report the serendipitous discovery of a new WR star that was
selected for study due to its photometric variability. Even weak variability can
mark unusual stellar types and will be an important method of identifying
rare stars in the era of the Large Synoptic Survey Telescope \citep{borne08}.

\begin{figure}[ht!]
\epsscale{.80}
\plotone{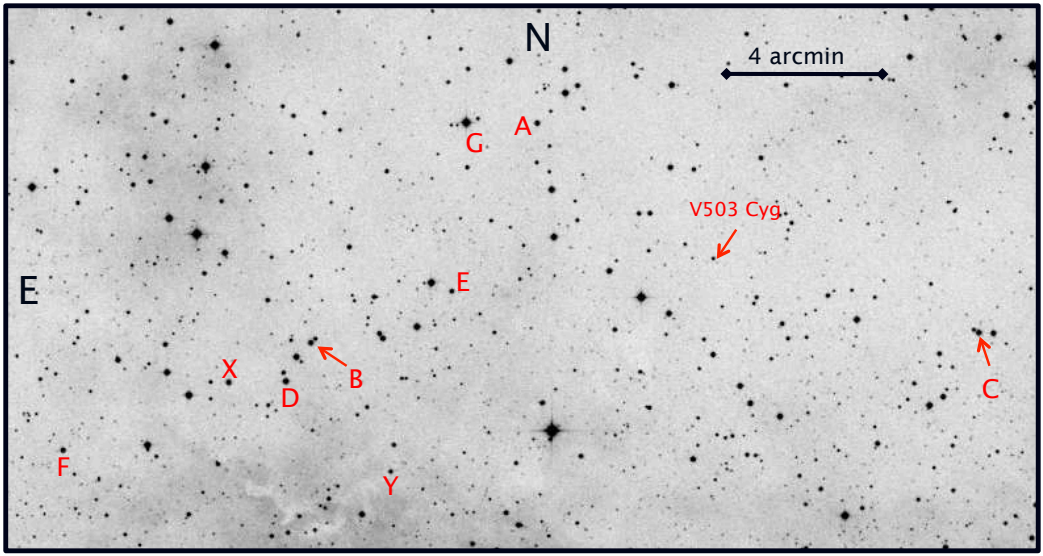}
\vspace{-6.0cm}
\caption{A red Digitized Sky Survey image around V503~Cyg showing the
variable stars (stars A-G) identified in this study. Star~X
is the comparison star used in the photometry and star~Y is a check
star. Star~B is a new Wolf-Rayet star, \name .
  \label{fig1}}
\end{figure}

\section{Observations}

\subsection{Photometry}

In July 2011, we obtained time-resolved photometry of a superoutburst
of cataclysmic variable V503 Cygni with a 28-cm aperture 
Schmidt-Cassegrain telescope atop Jordan Hall on the University of
Notre Dame campus. Seventeen nights of time-series data were taken using a commercial
SBIG CCD camera and are listed in Table~\ref{table1}. 

During data reduction, we noticed seven nearby field stars (stars A-G
in Figure~1) that appeared to vary in brightness on a time scale of hours.
Stars A, C, and E are Algol-type eclipsing binaries, and star F is a W UMa eclipsing
binary. Star D exhibited short-term pulsations consistent with a very
low-amplitude Delta Scuti type variable. Star G, a known Be star, showed
quasi-periodic pulsations which photometry from SuperWASP \citep{street04}
subsequently confirmed.

The very red Star~B showed weak, irregular variability. Star~B corresponds
to USNO-B1.0 1336-0379707 with R2=12.7~mag and a B2$-$R2 color of 3.0~mag. Its position
at $\alpha_{J2000}=$ 20:28:14.54, $\delta_{J2000}=$ +43:39:25.6 is within
6~arcsec of HBH$\alpha$~4203-27 in the Hamburg-Bergedorf H$\alpha$ emission line catalog
\citep{kohoutek99}. The catalog notes that the H$\alpha$ line of HBH$\alpha$~4203-27
is over-exposed on a strong continuum, but no further classification is given.

Additional unfiltered observations of Star~B by 
Elena Pavlenko, Maksim Andreev, and Aleksej Sosnovskij at the 
Crimean Astrophysical Observatory in Ukraine also showed variation 
(Pavlenko, private communication). However, relative photometry of stars
with extreme color differences can create the appearance of variability
due to differential extinction with airmass.  The suspected
variability prompted us to obtain a spectrum of the star so that we
could ascertain its nature.

\subsection{Spectroscopy}

We first obtained spectra of Star~B using the same 28-cm
Schmidt-Cassegrain telescope and CCD camera as for the photometry,
but added a `Star Analyser' (manufactured by Paton Hawksley
Education Ltd.)  100-line-per-millimeter grism. The
resulting spectrum has a dispersion of 23~\AA/pixel. Spectra
of Star~B were obtained over the course of three 
nights starting 2011 July 31 (UT). The dispersion direction was rotated
so that the target spectrum did 
not overlap any nearby stellar spectra or zero-order images.  As an 
additional safeguard, we obtained a deep, unfiltered image of the 
field to compare with the summed spectrum to verify the absence of
contamination from zero-order images of faint stars.

A spectrum of $\alpha$~Lyrae was also taken and the positions of
the Balmer absorption lines relative to the zero-order image
were used to calibrate the wavelength as a function of pixel
position and the spectrum provided the flux calibration.

\begin{figure}[ht!]
\epsscale{.70}
\plotone{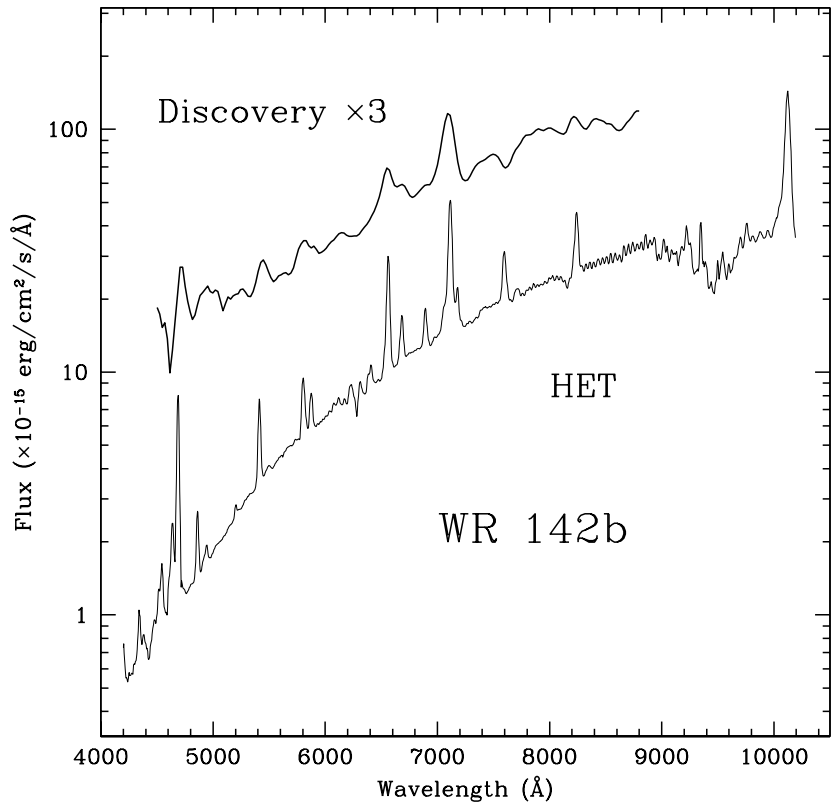}
\caption{The discovery spectrum of \name\ (shifted by a factor
of 3) compared with the HET spectrum confirming the
classification as a WR star. \label{fig2}}
\end{figure}

The total integration time for the spectrum of Star~B
was 167 minutes. Even in 1-minute subexposures, strong emission lines 
at approximately 6560~\AA\  and 7100~\AA\  were clearly visible.
The resulting spectrum is shown in Figure~\ref{fig2}.

The He~I line at 7064~\AA\ appears in many emission line objects and
could explain
the detection of a strong line at 7100~\AA , but this helium
line is rarely comparable in strength to H$\alpha$ at 6563~\AA. Thus,
we attributed the 7100~\AA\ feature to N~IV 7115~\AA ,
indicative of a nitrogen-rich WR star. Stars with strong
N~IV emission often show several He~II emission lines, making
the feature at 5400~\AA\ a good match to the 5411~\AA\
He~II line. Usually, the strongest He~II line in the optical is at
4686~\AA , but the combination of poor blue sensitivity and extremely
red continuum resulted in a poor signal-to-noise ratio shortward of 5000~\AA .

We then obtained a spectrum of the WR candidate using
the 9.2-m Hobby-Eberly Telescope. Three short (120-second) exposures were
taken and combined to avoid saturating the red end of the spectrum.
The HET spectrum, which offered
coverage from 4000~\AA\  to just beyond 1$\mu$m, confirmed 
the emission lines in our original data and revealed many more 
in both the blue and near-infrared portions of the star's
spectrum (see Figure~\ref{fig2}). In particular, we identify a large
number of He~II emission lines with the strongest at 
4686~\AA\ and 10124~\AA. Other lines and their
identifications are given in Table~\ref{table2}.
These firmly establish Star~B as a Wolf-Rayet star
of the WN class. Relying upon the WR nomenclature 
scheme set forth in van der Hucht (2001), we named the star
\name .

Once the nature of \name\ was firmly established by the HET spectrum,
we obtained Multi-Object Dual Spectrograph (MODS) data with
Large Binocular Telescope (LBT). MODS has sensitivity well into
the ultraviolet, which allowed us to search for a hot companion
as well as any unusual properties of the WR star. Eight individual
120-second exposures were taken with the 400 line/mm blue grating
on 2011 September 27 (UT). A 1.0-arcsecond wide slit was employed.

The bias was removed from the images and the data corrected for
variations in the flat field. The individual images were combined
before a one-dimensional spectrum was extracted. A line identified
as N~IV at 3478~\AA\ was clearly detected, showing that the sensitivity
of MODS extended shortward of 3500~\AA .

\section{Analysis}

\subsection{WR Classification}

\begin{figure}[ht!]
\epsscale{1.1}
\plottwo{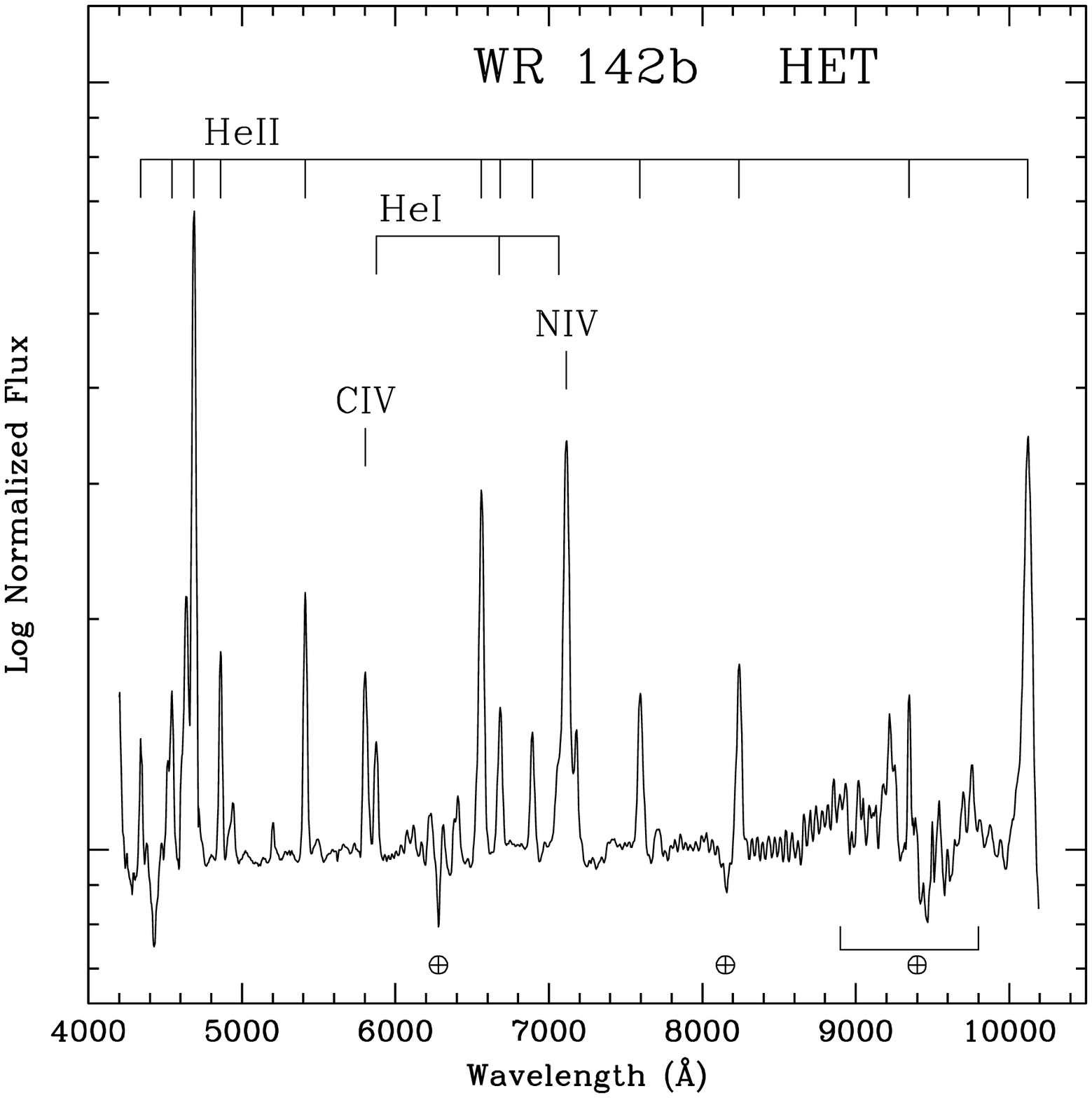}{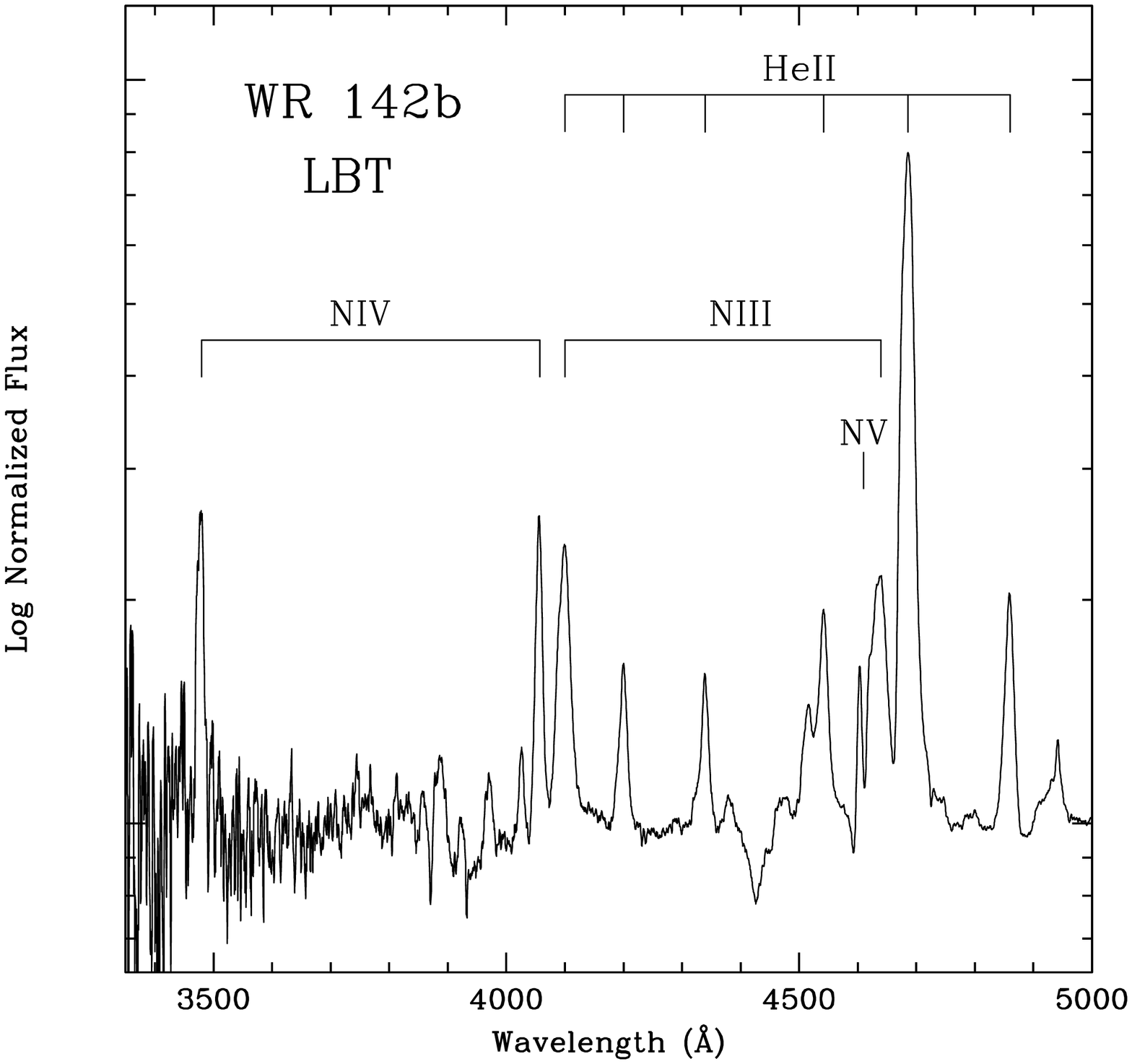}
\caption{{\bf Left:} The normalized HET spectrum of \name\ showing
the identifications of the major lines. The spectrum has been corrected
for telluric absorption but residual features are marked. {\bf Right:}
The blue end of the normalized LBT spectrum showing no evidence of a
hot companion star.  \label{fig3}}
\end{figure}

The strong N~IV, rich He~II spectrum, and overall lack of carbon emission
clearly make this a WN-type WR (Figure~\ref{fig3}). Subclassification of WN stars
is multi-dimensional, based on the \citet{smith96} system. The
degree of ionization is parametrized by the He~II 5411 to He~I 5875
line ratio. The observed equivalent width ratio is 2.6 for \name ,
placing it in the WN5 or WN6 class. The N~V to N~III ratio is
very small, consistent with a WN6 or WN7 star. The LBT spectrum
includes the N~IV line at 4057~\AA\  and its peak compared with the
4630~\AA\ blend is  consistent with
a WN5 or WN6 classification. The C~IV to He~II 5411
ratio is 0.7 and C~IV to He~I is 2.0, both at the WN5/6 boundary.
We conclude that \name\ is best matched by a WN6 ionization classification.

The He~II lines have widths of 24~\AA\ (FWHM) before accounting for
the instrumental resolution. This is within the range of normal line
widths, so \name\ should not be considered a broad-lined WR star.

Strong He~II emission can hide the presence of hydrogen in the spectrum
of WN stars because of the close coincidence of the Balmer lines
with the many Pickering series lines of He~II.
\citet{smith96} note that when hydrogen is present,
H$\beta$ and H$\gamma$ will increase the flux of the
corresponding Pickering lines while leaving the other Pickering
lines unperturbed. However, \name\ shows a monotonically
decreasing Pickering series, and we conclude that
it does not have detectable hydrogen in its atmosphere.

\subsection{Reddening by Dust}

The very red color of \name\ suggests a significant amount of
dust extinction local to the WR star or along the line-of-sight.
Emission line ratios \citep{conti90} and the slope of the continuum
\citep{morris93} have been used to estimate the reddening of individual WR stars.

The predicted spectrum of hydrogen recombination from a low-density nebula
is often compared with the observed emission to estimate extinction.
In contrast, the physical processes
of the atmospheres of helium-rich WR stars are complex and require
detailed modelling to accurately predict the line strengths. \citet{conti90}
have used an empirical approach and found that the ratio between the strong
UV He~II line at 1640~\AA\ is consistently a factor of 7.6 brighter than
the optical 4686~\AA\ line with only a 20\%\ scatter in 30 WR stars. This is very close
to the optically thin recombination ratio for these lines. Unfortunately, we
have not observed the 1640~\AA\ line, but we do have a measurement of
the bright IR line at 1.012~$\mu$m.

We have run the CLOUDY photoionization code \citep{ferland98} with enhanced helium
and over a range of temperatures and densities. We find that
when the 1640~\AA\ to 4686~\AA\ flux ratio is 8, the 4686~\AA\ to 10120~\AA\ 
flux ratio is between 5 and 6. The observed flux ratio is 0.031, meaning the
4686~\AA\ He~II line is between 160 and 200 times fainter than predicted. For a
\citet{ccm89} dust law, this corresponds to a reddening of $E(B-V)=2.15$
to 2.25 mag.

\citet{morris93} found that the continua of WN-type Wolf-Rayet stars between
the ultraviolet (UV) and near-infrared (NIR) wavelengths
are well matched by a power law with an index of $-2.72\pm 0.39$.
To create a continuum energy distribution (CED), we estimated the flux in
regions of the HET spectra with no significant emission
and we add the 2MASS NIR magnitude measurements ($J=8.769\pm 0.021$,
$H=7.861\pm 0.017$, $K=7.191\pm 0.020$) converted to flux. The optical and NIR
fluxes were obtained at different times for this variable star, but the
amplitude of the variation is small and should not dominate the error
in the slope estimate. The CED was
dereddened using the \citet{ccm89} law until the slope was best fit
by a power law with index $-2.72$. This slope is matched with a 
$E(B-V)=2.57\pm 0.14$ mag. The uncertainty in this extinction estimate comes directly from
the range of power law indices in the observed sample of WR stars. This
reddening estimate is more reliable than the line ratio technique, so we adopt
a reddening of $E(B-V)=2.6\pm 0.2$ mag for \name .

\subsection{Companion?}

Van der Hucht (2001) found that 39\%\ of known WR stars
harbor companions, usually of the OB type. The companions add
absorption lines at blue wavelengths and can generate
an UV bump in the continuum of the WR star. Our HET spectrum
does not reach far enough into the blue to test for these
effects, but we specifically obtained a LBT spectrum to
search for evidence of a companion.  The blue end of the LBT
spectrum is shown in Figure~\ref{fig3}. Narrow interstellar
absorption features are present, but there is no sign of stellar
absorption consistent with a companion, and the spectrum before continuum
normalization does not display a strong bump from a hot companion. We find
no evidence of a hot companion star to \name .

\subsection{Infrared Flux}

\begin{figure}[ht!]
\epsscale{.80}
\plotone{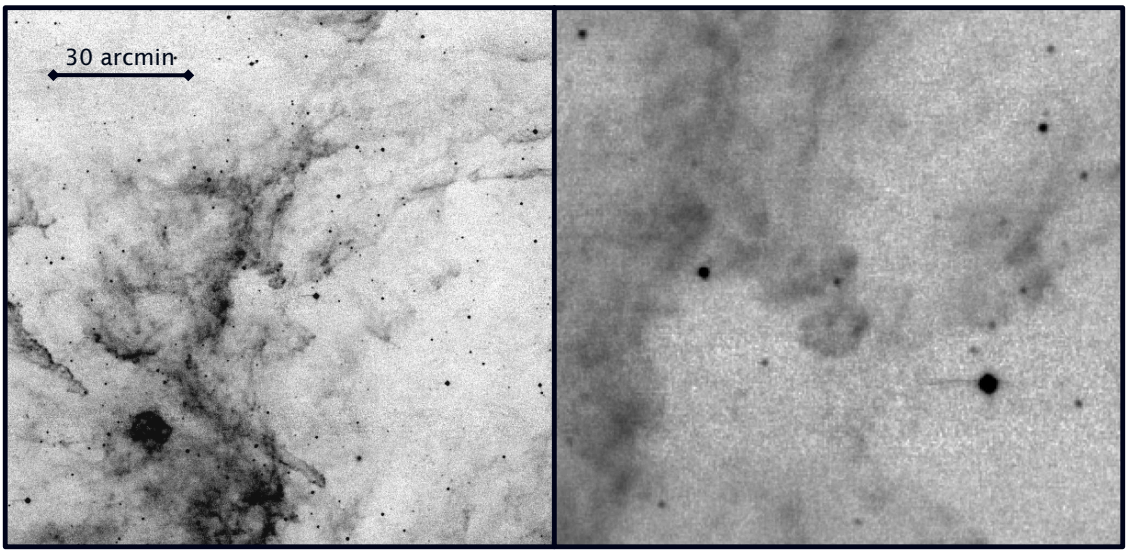}
\vspace{-6.0cm}
\caption{An 8.28~$\mu$m wavelength MSX image centered on \name . The left
panel shows a two degree region around the star. The right panel is a 0.5 degree
field showing \name\ in the center.   \label{farir}}
\end{figure}

The Cygnus region was observed with the Midcourse Space Experiment (MSX)
and the MSX6C point source catalog \citep{egan03} lists a source
within 0.6~arcsec of \name .
G081.5744+02.9135 has a measured flux of 0.346~Jy in band A (8.28 $\mu$m)
but is not detected in the longer-wavelength bands. The low
density of IR point sources in the survey combined with the strong positional
coincidence suggests that \name\ is detected at 8.28~$\mu$m. The MSX image of the field is
shown in Figure~\ref{farir} and a large amount of dust
emission is evident toward \name . In particular, a finger of
dust emission extends over the star and may be the
cause of the very high estimated extinction.


If we extrapolate the optical/near-IR flux after correcting for
our estimated reddening, we predict a flux of 3.1$\times 10^{-15}$
erg~cm$^{-2}$~s$^{-1}$~\AA$^{-1}$ at 8.28 $\mu$m. This is nearly a factor
of two larger than the MSX-observed flux after a 0.16-mag extinction
correction at 8~$\mu$m is applied. But given the large extrapolation and variability
of the star, the prediction is reasonably close.

If we fit a power-law slope to only the  2MASS magnitudes for \name\ and extrapolate
to 8.28~$\mu$m, we predict to a flux of 1.74$\times 10^{-15}$
erg~cm$^{-2}$~s$^{-1}$~\AA$^{-1}$, which matches the observed MSX flux.
The power-law index using only the NIR continuum is 3.2, which is slightly
steeper than when the optical is included. These extrapolations
imply no strong mid-IR emission from \name\ beyond that from
the stellar continuum. We conclude that there is no significant
circumstellar dust emission from \name\ and that the intense extinction
is extrinsic to the star.

\subsection{Variability}

In a study of northern WR stars, Moffat and Shara (1986)
found that at least half of the WRs in their survey display low-amplitude
(often 5\%) variability in optical wavelengths. As
noted in the photometry section, \name\ exhibited irregular, low-amplitude variation in 
our unfiltered observations. We subsequently reprocessed 
the photometry and used suitably red comparison and check star (stars X and Y in Table~\ref{fig1}),
selected by their $J-K$ colors. \footnote{The check star is simply another comparison
star which verifies that the primary comparison star is not
variable.} The weak variation seen in \name\  remained present in many
of the light curves, but its amplitude never exceeded 0.03 magnitudes.

\begin{figure}[ht!]
\epsscale{.70}
\plotone{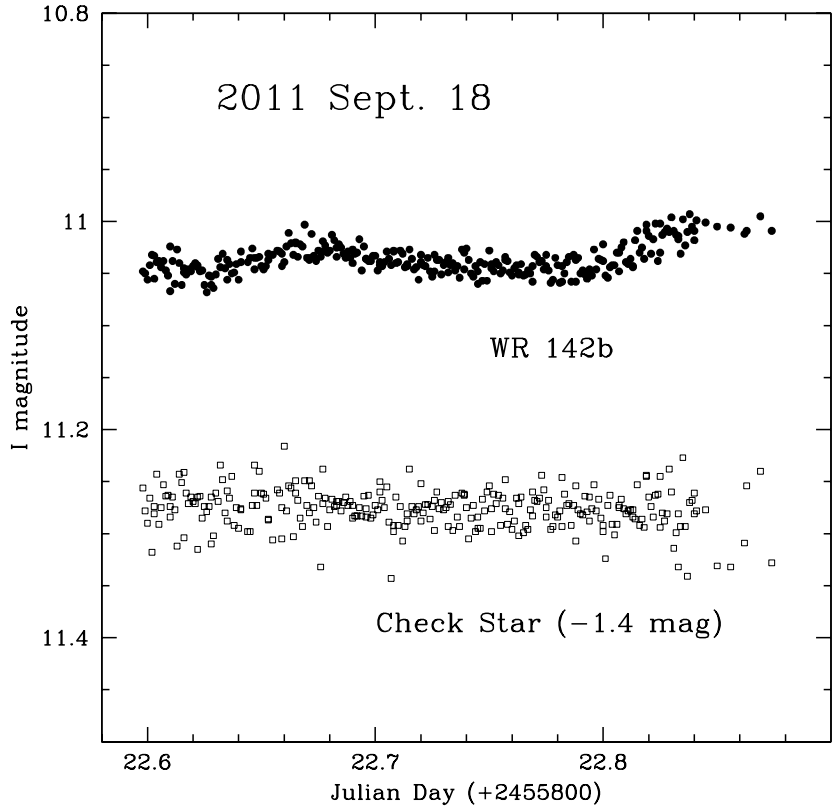}
\caption{A typical $I$-band light curve of \name . The magnitude
is calibrated assuming the comparison star has a brightness of
$I$=11.26 mag. The check star is shown shifted by $-1.4$ mag.   \label{lightcurve}}
\end{figure}

Since the data were unfiltered, we were concerned that the observed variability might be an artifact
caused by differential extinction. To minimize this effect, we obtained time-series photometry
on three nights using an $I$-band filter and red comparison and check stars.
All of the filtered light curves show low-amplitude
variability similar to that observed in the unfiltered data. The variability typically appears as 
a gradual brightening, followed by a slower fade, but no consistent period is apparent. Crucially, it 
is independent of airmass, and the check star never displays similar activity. Figure~\ref{lightcurve}
presents a representative photometric time series.


\name\ also shows night-to-night variability. The average $I$-band magnitude of the star on six nights over
a two-month span is given in Table~\ref{phot}. In particular, the infrared magnitude on 2011 September 12 (UT)
was 0.08~mag higher than it had been two days earlier, and two days later, it had faded by
0.11~mag.
The root-mean-squared (RMS) variation of \name\ over the six nights was 0.040~mag while the check star,
which was four times fainter, varied by only 0.009~mag. We conclude that the weak variability
suspected in the unfiltered data is confirmed with the better-controlled $I$-band photometry.


\section{Conclusion}

	We have serendipitously discovered a Wolf-Rayet star, \name,
by first detecting its photometric variability and then obtaining
spectroscopy. A HET spectrum shows emission lines of
He~I, He~II, N~IV and C~IV. After
examining the strengths of \name\ spectral lines, we
find that it is best classified as a WN6 with no detectable hydrogen
and normal line widths. The spectrum shows good evidence
for a significant reddening due to dust. We estimate the
reddening is between $E(B-V)=2.2$ and 2.6 mag and most likely toward
the higher value. For a standard dust
law the star is dimmed by 8 mag in the $V$ band and would have 
a visual brightness of $V_0=6.6$ mag with no extinction. However,
the total-to-selective extinction ratio can vary for strongly extincted
WR stars \citep{turner11} and $R_V$ could be as high as 4, making \name\ a
naked-eye object if not for the dust. We do not detect a hot companion
despite obtaining LBT spectra sensitive down to 3400~\AA .


\acknowledgments

We wish to thank Perry Berlind of the Fred Whipple Observatory, Sergey Rostopchin
of McDonald Observatory and Rick Pogge of The Ohio State University.
Additionally, Elena Pavlenko, Maksim Andreev, and Aleksej 
Sosnovskij at the Crimean Astrophysical Observatory and Brian Skiff of the Lowell Observatory
kindly contributed photometry of \name . JCW is supported in part by NSF grant
AST-1109801.

The LBT is an international collaboration among institutions in the United States,
Italy and Germany. LBT Corporation partners are: The University of Arizona on behalf
of the Arizona university system; Istituto Nazionale di Astrofisica, Italy; LBT
Beteiligungsgesellschaft, Germany, representing the Max-Planck Society, the
Astrophysical Institute Potsdam, and Heidelberg University; The Ohio State University,
and The Research Corporation, on behalf of The University of Notre Dame,
University of Minnesota and University of Virginia.

\clearpage

\begin{deluxetable}{lccc}
\tablewidth{0pt}
\tablecaption{Time-Resolved Photometry Log \label{table1}}
\tablehead{
\colhead{UT Date} & 
\colhead{UT Start} &
\colhead{UT End} &
\colhead{Filter}  \\
\colhead{(2011)} &
\colhead{(hours)} &
\colhead{(hours)} &
\colhead{ } \\ }
\startdata
Jul 05    &  3.85 & 8.92 & clear  \\
Jul 08    &  5.02 & 8.59 & clear  \\
Jul 12    &  4.46 & 7.88 & clear  \\
Jul 14    &  5.38 & 9.43 & clear  \\
Jul 16    &  2.65 & 8.16 & clear  \\
Jul 17    &  2.57 & 5.12 & clear  \\
Jul 20    &  2.38 & 6.29 & clear  \\
Jul 21    &  4.21 & 9.12 & clear  \\
Jul 25    &  2.58 & 7.68 & clear  \\
Jul 27    &  2.53 & 5.32 & clear  \\
Jul 30    &  2.84 & 9.16 & clear  \\
Jul 31    &  2.19 & 4.07 & clear  \\
Aug 01    &  2.04 & 8.25 & clear  \\
Aug 02    &  2.74 & 4.65 & clear  \\
Aug 05    &  2.46 & 8.23 & $I$-band \\
Sep 12    &  4.92 & 8.81 & $I$-band \\
Sep 18    &  2.36 & 8.98 & $I$-band \\
\enddata
\end{deluxetable}

\begin{deluxetable}{lcccc}
\tablewidth{0pt}
\tablecaption{Emission Lines \label{table2}}
\tablehead{
\colhead{Observed} &
\colhead{Equ. Width} &
\colhead{Flux} &
\colhead{ID}   &
\colhead{Rest}   \\
\colhead{Wavelength (\AA)} &
\colhead{(\AA)} &
\colhead{($10^{-14}$ erg/cm$^2$/s)} &
\colhead{ } &
\colhead{Wavelength (\AA)} \\ }
\tiny
\startdata
3478.6 & 19.2 &  0.12 & N~IV  & 3478 \\
3887.7 & 5.3  &  0.10 & He~I  & 3888  \\
3970.0 & 4.1  &  0.09 & He~II & 3968  \\
4025.8 & 3.8  &  0.10 & He~II & 4025  \\
4056.0 & 20.0 &  0.59 & N~IV  & 4058 \\
4098.5 & 26.9 &  0.90 & N~III+He~II & Blend \\
4199.7 & 9.8  &  0.40 & N~III+He~II & Blend \\
4341.8 & 9.6  &  0.68 & He~II & 4339  \\
4516.2 & 6.1  &  0.56 & N~III & Blend  \\
4543.1 & 14.8 &  1.5  & He~II & 4542  \\
4605.2 & 5.1  & 0.55  & N~V & Blend  \\
4637.0 & 25.8 &  3.3  & N~III & Blend  \\
4687.6 & 120 & 16.6  & He~II & 4686  \\
4861.3 & 19.7 &  2.8  & He~II & 4859.3  \\
4943.0 & 2.7  &  0.47  & N~V & 4940  \\
5201.2 & 1.9  &  0.45  & N~IV\tablenotemark{a} & 5204  \\
5412.6 & 31.4 & 11.1  & He~II & 5411  \\
5804.1 & 23.5 & 12.8  & C~IV & 5808  \\
5875.0 & 11.9 &  6.6  & He~I & 5875  \\
5931.6 & 0.28 & 0.17  & He~II (Pf25) & 5932  \\
5955.7 & 0.22 & 0.14  & He~II (Pf24) & 5953  \\
5979.6 & 0.21 & 0.13  & He~II (Pf23) & 5977  \\
\tableline
\tablebreak
6004.3 & 0.27 & 0.18  & He~II (Pf22) & 6004  \\
6037.9 & 0.76 & 0.51  & He~II (Pf21) & 6037  \\
6075.7 & 1.1  & 0.78  & He~II (Pf20) & 6074  \\
6119.7 & 1.6  & 1.2   & He~II (Pf19) & 6118  \\
6171.7 & 1.3  & 0.92  & He~II (Pf18) & 6171  \\
6234.6 & 5.4  &  3.8  & He~II (Pf17) & Blend  \\
6313.2 & 3.5  &  2.8  & He~II (Pf15 & 6312  \\
6380.4 & 2.5  &  2.1  & ?    & ...   \\
6406.9 & 7.0  &  6.1  & He~II (Pf15) & 6406  \\
6561.2 & 67.7 & 65.7  & He~II & 6560  \\
6665.2 &  2.7 &  2.8  & He~I & Blend  \\
6684.6 & 14.9 & 16.6  & He~II (Pf13)  & 6683   \\
6893.2 & 13.0 & 17.2  & He~II (Pf12)   & 6890   \\
7062.5 & 12.8 & 19.2  & He~I  & 7064  \\
7114.7 & 97.9 & 147   & N~IV  & 7110   \\
7180.1 & 15.1 & 20.6  & He~II (Pf11) & 7178   \\
7596.5 & 22.3 & 44.2  & He~II (Pf10) & 7593  \\
8240.3 & 27.1 & 71.5  & He~II (Pf9) & 8237  \\
9348.3 & 11.8 & 31.8  & He~II (Pf8) & 9345  \\
10122  & 118.2 & 527   & He~II & 10120  \\
\enddata
\tablenotetext{a}{Possible identification based on \citet{ralchenko11}}
\end{deluxetable}

\begin{deluxetable}{lccc}
\tablewidth{0pt}
\tablecaption{Long-Term Variability of \name \label{phot}}
\tablehead{
\colhead{UT Date} & 
\colhead{Julian Day} &
\colhead{Average $I$$^a$} &
\colhead{Check Star $I$$^a$} \\
\colhead{(2011)} &
\colhead{(+2455000)} &
\colhead{(mag)} &
\colhead{(mag)} \\ }
\startdata
Jul 10    &  752.8113 & $10.983\pm 0.002$ & $12.674\pm 0.004$   \\
Aug 05    &  778.7229 & $10.995\pm 0.001$ & $12.692\pm 0.003$   \\
Sep 12    &  816.7735 & $11.013\pm 0.001$ & $12.672\pm 0.003$   \\
Sep 14    &  818.6175 & $10.930\pm 0.002$ & $12.691\pm 0.006$   \\
Sep 16    &  820.6495 & $11.045\pm 0.002$ & $12.696\pm 0.006$   \\
Sep 18    &  822.7304 & $11.045\pm 0.002$ & $12.680\pm 0.002$   \\
Sep 22    &  826.6625 & $10.982\pm 0.002$ & $12.680\pm 0.006$   \\
\hline
Mean$\pm$RMS & & 10.999$\pm 0.040$ & 12.684$\pm 0.009$ \\
\enddata
\tablenotetext{a}{Based on the USNO-B1.0 magnitude for the comparison star
of $I$=11.26 mag. The error estimate includes only Poisson noise.}
\end{deluxetable}




\end{document}